\documentclass[prl,aps,epsfig,twocolumn]{revtex4-2}
\usepackage{amsfonts,amssymb,amscd,amsthm}
\usepackage{graphicx}
\usepackage{mathrsfs}
\usepackage[intlimits]{amsmath}
\usepackage[colorlinks, citecolor=red,hyperfootnotes=false]{hyperref}
\usepackage{physics}
\usepackage{lineno}

\usepackage{natbib}
\usepackage{footnote}

\begin{document}
\title{hqQUBO: A Hybrid-querying Quantum Optimization Model Validated \\ with 16-qubits on an Ion Trap Quantum Computer for Life Science Applications}
\author{Rong Chen\footnote{These authors contribute equally to this work}$^{\dag1}$, Quan-Xin Mei$^{\dag2}$, Wen-Ding Zhao$^{2}$, Lin Yao$^{2}$, Hao-Xiang Yang$^{2}$\\ Shun-Yao Zhang$^{*1}$, Jiao Chen$^{*1}$, Hong-Lin Li\footnote{Corresponding author: hlli@ecust.edu.cn, chenjiao@lglab.ac.cn, shyzhang@lglab.ac.cn}$^{*1,3}$}

\affiliation{$^{1}$Lingang Laboratory, Shanghai, China}
\affiliation{$^{2}$HYQ Co., Ltd., Beijing, China}
\affiliation{$^{3}$Innovation Center for AI and Drug Discovery, School of Pharmacy, East China Normal University, Shanghai 200062, China}
\affiliation{$^{\dag}$These authors contributed equally to this work}
\affiliation{$^{*}$Corresponding authors. E-mail: shyzhang@lglab.ac.cn; chenjiao@lglab.ac.cn; hlli@lglab.ac.cn}

\begin{abstract}
AlphaFold has achieved groundbreaking advancements in protein structure prediction, exerting profound influence across biology, medicine, and drug discovery. However, its reliance on multiple sequence alignment (MSA) is inherently time-consuming due to the NP-hard nature of constructing MSAs. Quantum computing emerges as a promising alternative, compared to classical computers, offering the potentials for exponential speedup and improved accuracy on such complex optimization challenges. This work bridges the gap between quantum computing and MSA task efficiently and successfully, where we compared classical and quantum computational scaling as the number of qubits increases,  and assessed the role of quantum entanglement in model performance. Furthermore, we proposed an innovative hybrid query encoding approach hyQUBO to avoid redundancy, and thereby the quantum resources  significantly reduced to a scaling of $\mathcal{O}(NL)$. Additionally, coupling of VQE and the quenched CVaR scheme was utilized to enhance the robustness and convergence. The integration of multiple strategies facilitates the robust deployment of the quantum algorithm from idealized simulators (on CPU and GPU) to real-world, noisy quantum devices (HYQ-A37). To the best of our knowledge, our work represented the largest-scale implementation of digital simulation using up to 16 qubits on a trapped-ion quantum computer for life science problem, which achieved state of the art performance in both simulation and experimental results. Our work paves the way towards large-scale simulations of life science tasks on real quantum processors.
\end{abstract}
\maketitle

\section{Introduction}
The protein folding problem has been a central question in biology for over 50 years. Understanding protein structures is critical for deciphering its function, interaction and role in biological processes. AlphaFold\cite{jumper2021highly} achieved unprecedented accuracy in predicting protein structures, comparable to experimental techniques such as X-ray crystallography\cite{wayment2024predicting}. A foundational component of AlphaFold's success is multiple sequence alignment (MSA), which captures the co-evolutionary patterns between amino acid residues in homologous proteins to ensure accurate predictions. However, constructing MSAs, particularly for large or diverse protein families, is computationally intensive and time-consuming.\\

The complexity of aligning sequence increases exponentially as the number of sequences and their lengths grow.  Larger alignment tasks require more MSAs and thus greater computational power. The most accurate MSA algorithm is based on the dynamic programming approach\cite{needleman1970general}, which assigns scores to each type of mutation, aiming to find the alignment with the lowest sum of pairwise alignment scores (SP).  Solving the MSA with SP scores is proved to be NP-hard\cite{wang1994complexity} and its computational demand rapidly grows with the number $N$ of sequences, and the length $L$ as $O(L^N)$\cite{corpet1988multiple}.  Consequently, exact sequence alignment is limited to pairwise alignment, missing out on deeper evolutionary insights that arise from aligning multiple sequences.  Alternative heuristic methods, such as progressive alignment\cite{notredame2000t} and profile HMMs alignment\cite{finn2011hmmer}, are typically employed for MSA to address these computational challenges, albeit at the expense of optimality.  With the exponential expansion of sequence datasets and the increasing demand for precision in protein structure prediction, there is an increasing need for more efficient methods for multi-sequence similarity analysis.\\

Quantum computing represents a groundbreaking paradigm, harnessing quantum-mechanical effects such as superposition, interference, and entanglement to address problems that classical computers struggle with. Significant advancements in both hardware and algorithms have been made in the past decade, including demonstrations of quantum advantages in quantum simulations of Ising models\cite{guo2024site}, stochastic circuit simulations\cite{arute2019quantum}, and boson sampling problems\cite{jiuzhang}. However, we are still in the Noisy Intermediate-Scale Quantum (NISQ) era, where quantum devices, such as superconducting circuits, trapped ions, and neutral-atom systems, have a limited number of imperfect qubits, restricting the width and depth of quantum circuits.  Recently,  Google's Willow\cite{acharya2024quantum} quantum computer surpassed the quantum error correction threshold,  signaling progress toward fault-tolerant quantum computing.  Nevertheless, achieving quantum advantage for real-world applications remains a formidable challenge. Therefore, developing efficient, domain-specific quantum algorithms and benchmarking them on current quantum devices is crucial to fully exploit the potential of quantum computing.  \\
 
Among the most promising application areas are the life science, particularly bioinformatics and pharmacology.  In bioinformatics, A.S. Boev \textit{et al.}\cite{boev2021genome} demonstrated quantum techniques for genome assembly,  enhancing computational efficiency.  Perdomo-Ortiz \textit{et al.}\cite{perdomo2012finding} proposed a lattice-based model for simulating protein folding via quantum annealing,  and later Robert developed a variational quantum algorithm,  demonstrating the folding of a 7-amino-acid peptide on IBM's quantum computer\cite{robert2021resource}.  In pharmacology,  quantum algorithms have shown promise for drug design.   Kirsopp \textit{et al.}\cite{kirsopp2022quantum} have quantified the protein-ligand interactions using the latest superconducting transmon (IBM) and ion-trap (Honeywell) quantum devices for the first time.  Zha \textit{et al.}\cite{zha2023encoding} proposed a pose sampling method,  achieving sampling speeds 1000 times faster than Glide SP.  Additionally,  quantum circuits integrated with classical models have been used to generate molecules with improved physicochemical properties\cite{kao2023exploring} and design new KRAS inhibitors in cancer therapy\cite{vakili2024quantum}.\\

\begin{figure*}[tbp]
    \includegraphics[width=0.9\linewidth]{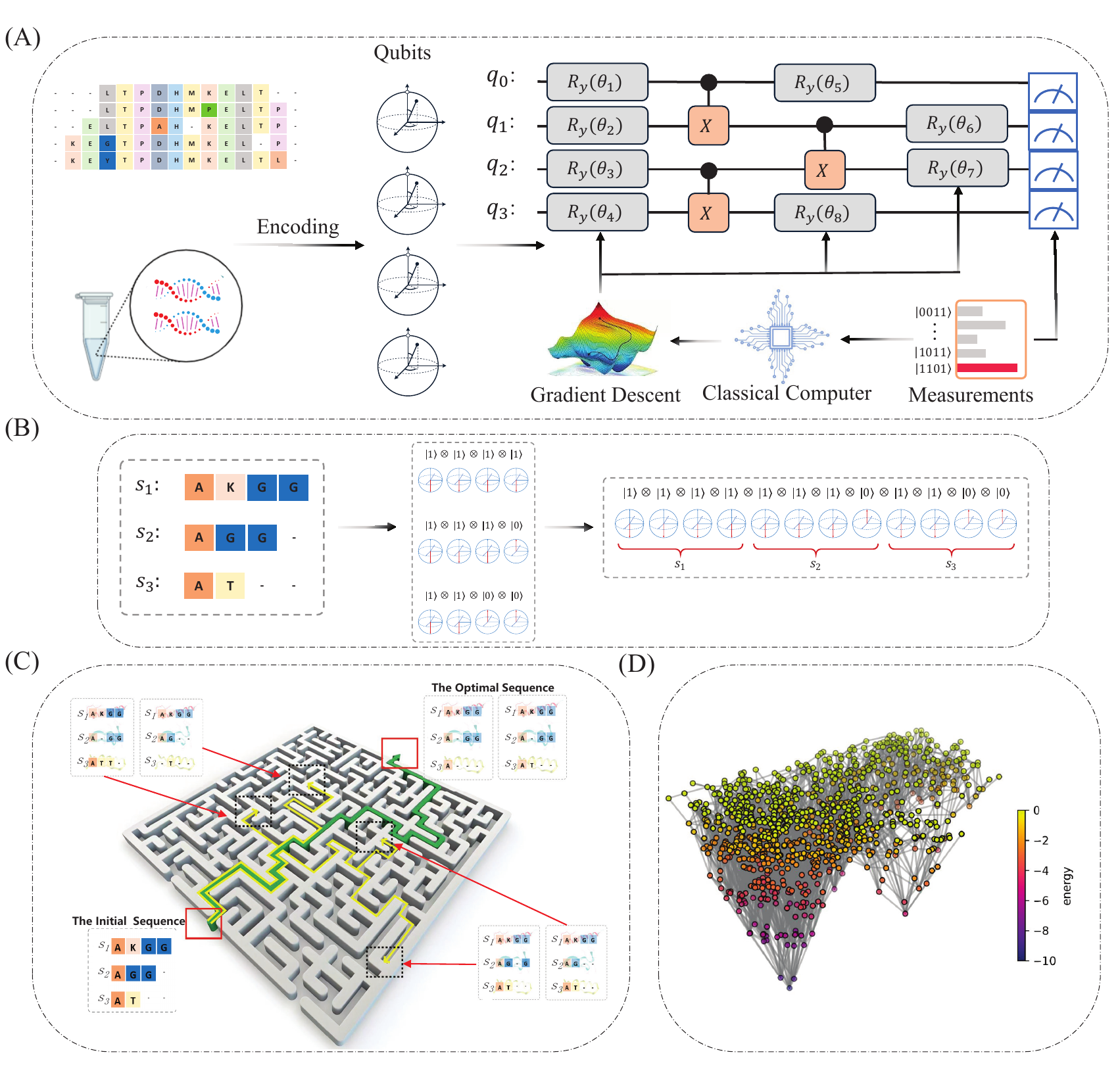}
    \caption{Schematic of quantum algorithms for solving the MSA problem.\textbf{(A)} The flowchart of a quantum variational algorithm used to solve the MSA problem, which involves finding the optimal alignment of DNA or protein sequences by inserting spaces to maximize similarities. The MSA problem is encoded into a hqQUBO model by mapping sequences to quantum states. Subsequently, the reformulated problem is addressed using a quantum variational algorithm, a quantum-classical hybrid approach comprising variational quantum circuit operations, quantum state measurements, and gradient computations performed on a classical computer. \textbf{(B)}  An example illustrating the encoding methods for this MSA instance. Three gene sequences, $s_{1}$, $s_{2}$ and $s_{3}$, are encoded into quantum states $|qs_{1}\rangle$, $|qs_{2}\rangle$ and $|qs_{3}\rangle$, respectively. The spare bits are represented by the quantum state $|0\rangle$. Finally, these are integrated into a complete quantum state $|\psi\rangle$ according to the order of sequences. \textbf{(C)}  Different optimization trajectories of MSAs is visualized as a maze,  where quantum algorithm has the ability to superpose all possible paths starting from the initial sequence,  simultaneously exploring different paths and identifying the optimal sequence alignment.  The green path indicates the optimal MSA,  while the yellow paths represent some local miminas. \textbf{(D)} The low-energy landscape of MSA states for the problem in Fig.\ref{fig:4} (D)  is plotted, where MSA states are represented as nodes, and adjacent nodes with a Hamming distance of 1 are connected by edges. The color bar and vertical position distinguish the energy levels of MSA states.  The two salient vertices indicate the global and local minimas, respectively.}
    \label{fig:1}
\end{figure*}

In this study, we formulated the MSA problem as an innovative hybrid-querying quadratic unconstrained binary optimization problem (dubbed as hqQUBO model) (see Fig.\ref{fig:1}). Our method employs a hybrid quantum-classical algorithm comprising two components: a sampling module executed on quantum devices and a query module processed on classical computers. This design significantly reduces the quantum resource requirements for the MSA problem to $\mathcal{O}(NL)$, thereby enhancing compatibility with the limited resources of current quantum devices in the NISQ era.  We conducted experiments on both a classical computer and a trapped-ion quantum computer (HYQ-A37) provided by HYQ Co., Ltd.  The HYQ-A37 features 37 high-quality qubits,  each of which can be individually manipulated via acousto-optic deflectors. We will first present the simulation and experimental results, followed by a detailed description of the algorithm optimized for efficient MSA implementation.  Our simulations and experiment on an actual ion-trap quantum processor demonstrated the potential of quantum algorithms in MSA tasks. To the best of our knowledge, this work represented the first digital simulation utilizing up to 16 qubits on a trapped-ion quantum computer, specifically applied to life science tasks.

\section{Results}\label{sec2}
Encoding real-world applications into quantum models naturally involves two steps: first, mapping the application to a QUBO (Quadratic Unconstrained Binary Optimization) problem and then formulating the QUBO as an Ising model, with the solution corresponding to the ground state of the model.  The QUBO framework is widely utilized in various combinatorial optimization problems, serving as a bridge between classical and quantum computational technologies\cite{glover2022quantum}. Once the QUBO model is derived from a real-world application, the next step is quantum optimization,  which can be achieved through variational quantum algorithms or quantum annealing.\\

In this work, we encoded MSAs into quantum states (referred to as MSA states), as illustrated in Fig.\ref{fig:1}(B). The aligned $N$ sequences, each with a maximal length $L$,  are concatenated into a $NL$-qubit state (or MSA state),  where the state $|1\rangle$ indicates the presence of an amino acid residue while the state $|0\rangle$ refers to a gap. The aligned sequences can then be scored according to Eq.\ref{eq:score}  in the methods section.  Thus,  given a MSA state,  the score (or energy) can be uniquely determined.  In our algorithm, the sampling module utilizes a hardware-efficient variational quantum circuit executed on a quantum computer to generate the MSA states, while the query module scores these states on a classical computer, as depicted in the flowchart in Fig. \ref{fig:1}(A).\\

We observed that solving the MSA problem is analogous to finding a path through a maze (illustrated in Fig.\ref{fig:1}(C)). In traversing the maze, one may encounter dead ends that correspond to local minima of the MSA problem.  To illustrate this concept more intuitively, we plotted the energy landscape of the MSA states, with edges connecting pairs of states differing by a Hamming distance of 1, as depicted in Fig.\ref{fig:1}(D). This landscape clearly reveals both the global and local minima, and the trajectories along the edges correspond to paths in the maze. Traditional classical algorithms explore one path at a time and are prone to getting stuck in local minimas.  In contrast,  quantum algorithms have the potential to simultaneously explore all paths by leveraging the superposition of the MSA states to efficiently search for the optimal solution.\\

\begin{figure*}[tbp]
    \includegraphics[width=\textwidth]{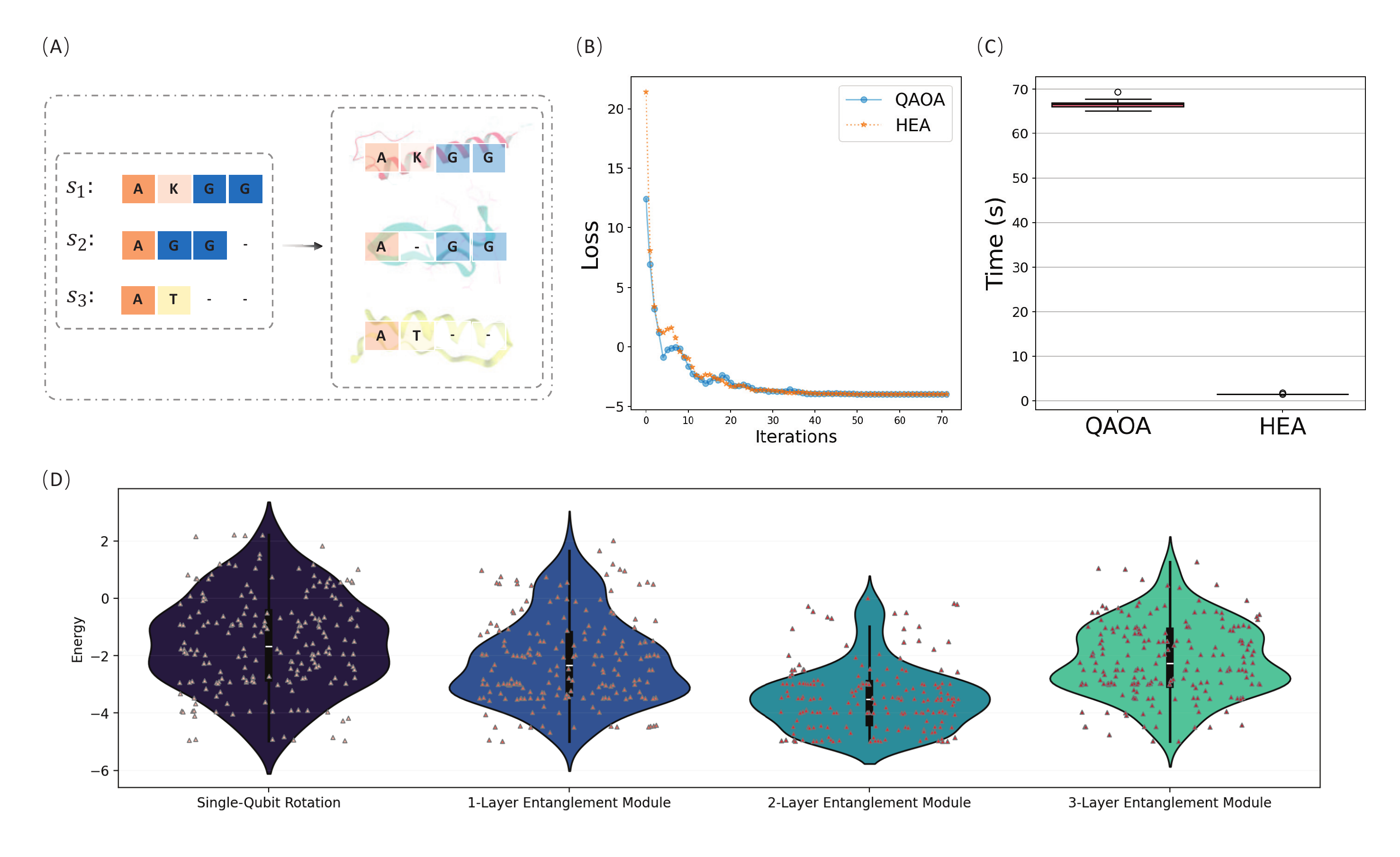}
    \caption{\textbf{Comparison between QAOA Ansatz and the Hardware Efficient Ansatz.} \textbf{(A)} A simple example with 12 qubits. \textbf{(B)} Convergence optimized by QAOA and HEA, respectively. \textbf{(C)} Box plot of time required for QAOA vs HEA to perform a training session. Note that these simulations were carried on a GPU processor. \textbf{(D)} The violin plot of the optimized energy distribution for hardware-efficient circuit structures.}
    \label{fig:2}
\end{figure*}

Based on the MSA encoding, we developed a hqQUBO model (see Methods for details) and proposed a variational Hybrid Quantum-Classical approach for the MSA problem, which significantly reduces the scaling of quantum resources to $\mathcal{O}(NL)$. In contrast, the Quantum Approximate Optimization Algorithm (QAOA), a polynomial-time scheme derived from the Trotterization of an infinite-time adiabatic algorithm \cite{sarkar2021quaser}—typically operates on a QUBO model, requiring at least $\mathcal{O}(N^2L)$ qubits for the MSA problem \cite{madsen2023multi}. To clarify the fundamental differences between QAOA and HEA, we presented a concise example in Fig.\ref{fig:2}. In this demonstration, 12 qubits were utilized, as shown in Fig.\ref{fig:2}(A). In particular, Fig.\ref{fig:2}(B) reveals that HEA attains convergence values comparable to those of QAOA.  Moreover, Fig.\ref{fig:2}(C) highlights the substantial advantage of HEA in time efficiency, with its required time cost significantly lower than that of QAOA. This discrepancy is primarily attributed to the large difference in the depths of their respective quantum circuits. In this case, the QAOA ansatz reaches a depth of 4486 layers, while the HEA circuit has only 7 layers. As the number of qubits grows, the computational cost of QAOA becomes increasingly prohibitive, further highlighting the superiority of HEA in terms of scalability and efficiency.\\


\subsection{Entanglement and Quenched CVaR Scheme}
Quantum entanglement is a fundamental resource in quantum computing. In principle, quantum entanglement expands the subspace beyond product states and induces quantum correlations among different sequences. To demonstrate how quantum entanglement improves the optimization of the MSA problem, we conducted numerical simulations employing various hardware-efficient quantum circuit architectures with differing levels of entanglement, as shown in Fig.\ref{fig:2}(D).\\

We benchmarked the quantum circuit featuring a single-qubit rotation layer against a hardware-efficient ansatz (HEA) comprising one, two, and three entanglement layers (see Appendix A). The experiment was conducted 200 times with random initial parameters on a Graphics Processing Unit (GPU) simulator, and the final energy was measured after 100 iterations. As indicated by the violin plots in Fig.\ref{fig:2}(D), the average energy associated with the 1-layer entanglement module is much lower than that of the circuit with only single-qubit gates. As the degree of entanglement increases, especially in the 2-layer entaglement HEA, we observed a marked increase in the number of samples converging to the ground state. However, when the entanglement is further increased, the performance begins to decline, which is likely due to overfitting. This trend highlights the critical role of modest quantum entanglement in solving MSA problems. Given that the optimized result typically constitutes a superposition of numerous MSA states, a decrease in energy indicates a greater contribution from low-energy MSA states. Furthermore, when multiple optimal solutions are available, quantum entanglement facilitates the simultaneous superposition of all optimal MSA states\\

Despite quantum entanglement, we found that Conditional Value-at-Risk (CVaR), defined as a tail-distribution-based loss function\cite{barkoutsos2020improving}, also contributes to the optimization of the MSA problem. Building upon the CVaR framework, we developed a two-stage optimization scheme that has been demonstrated to reduce the number of optimization steps and mitigate the risk of becoming trapped in local minima, as evidenced by our numerical simulations (see Methods for details).

\subsection{Simulations on CPU \& GPU Processors}
We conducted MSA simulations on a standard workstation equipped with Central Processing Unit (CPU) and GPU, implementing noise-induced and noiseless models for comparison. In the GPU (\textcolor{blue}{NVIDIA GeForce RTX 3060}) simulation, we designed two scenarios: 1) MSA with a reference sequence, 2) MSA without a reference sequence. Both scenarios were solved using a HEA circuit that employed 20 qubits and two layers (see Appendix A for details). For gradient computation, automatic differentiation and the Adam optimizer were utilized.\\    

\begin{figure*}[tbp]
    \includegraphics[width=\textwidth]{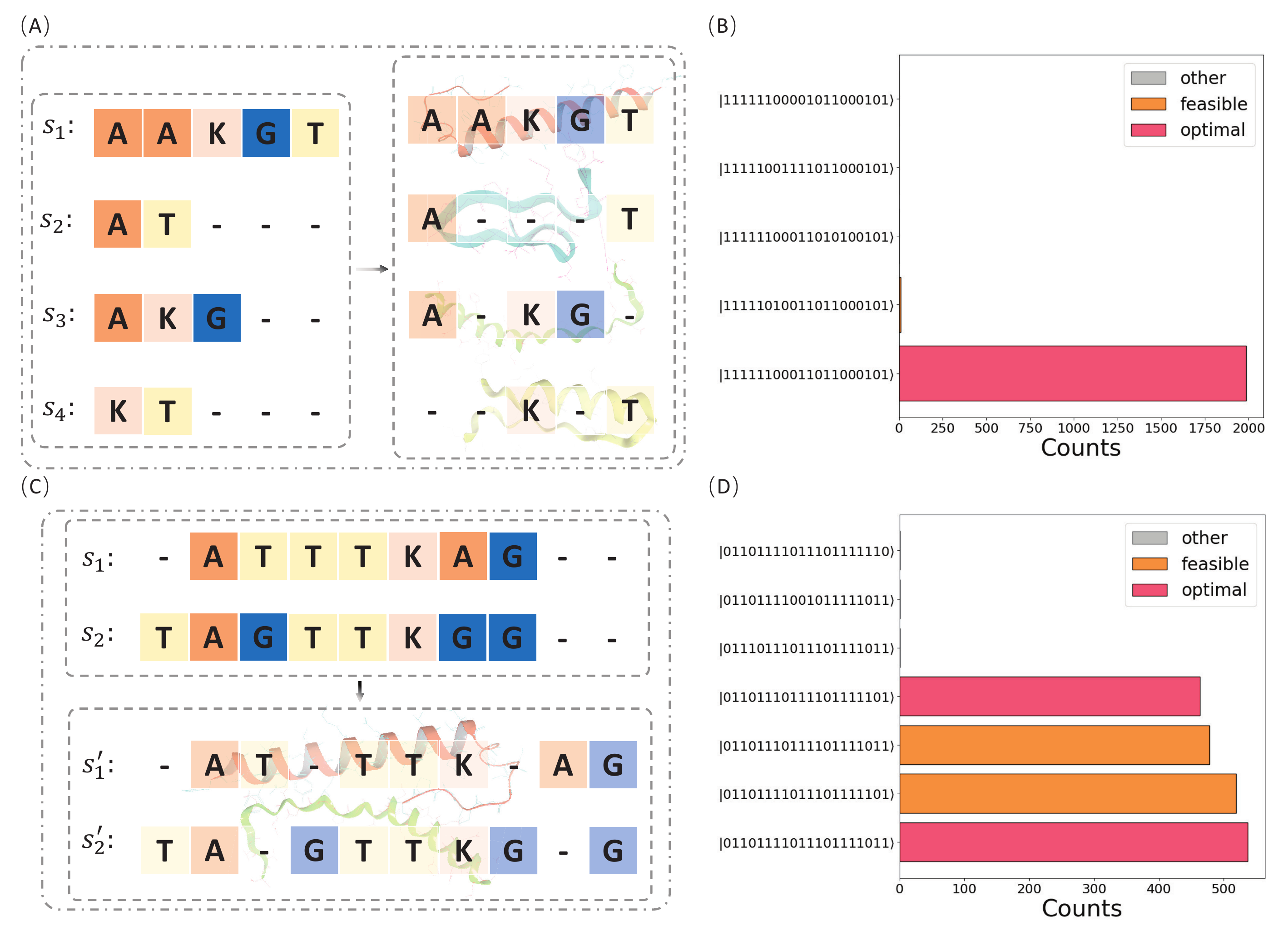}
    \caption{\textbf{The case with and without a reference sequence.} \textbf{(A)} An example with a reference sequence $s_{1}$, and a possible optimal alignment. The red part shows the reason why there are several optimal alignments. \textbf{(B)} The result based on HEA for (A), and the corresponding solution is $\ket{11111100011011000101}$. The feasible solution means that the number of letters remains constant, but the optimal similarity is not reached. \textbf{(C)} An example without a reference sequence, and a possible optimal alignment. \textbf{(D)} The result based on HEA for (C), the red bars mean the optimal solution, and the orange bars mean the feasible solutions. In fact, there is the only one-letter match error between the feasible solution and the optimal solution.}
    \label{fig:3}
\end{figure*}

For the first scenario, we designed a reference sequence $s_{1} = AAKGT$, and it can be matched with $s_2$, $s_3$, $s_4$ (see Fig.\ref{fig:3}(A)), where the score of the optimal similarity is $-10$. Figure \ref{fig:3}(B) shows the histogram results of the highest probability distribution from a 2000 shot simulation. From the graph, the red bar represents the optimal solution, denoted by the quantum state $\ket{11111100011011000101}$. The corresponding sequence alignment is given by: $AAKGT$, $A\_\_\_T$, $A\_KG\_$ and $\_\_K\_T$. Obviously, our model achieved the optimal solution and expected result.\\

As an extension of the previous task, we randomly generated two longer sequences without a reference, thereby providing more flexibility for the MSA task. The DNA fragments $s_1$ and $s_2$ have lengths of 7 and 8 bases, respectively, and both derived from a common ancestor sequence $s_0$, as shown in Fig.\ref{fig:3}(C). Determining the exact number of qubits required usually presents a challenge, as the length of the parent sequence and the optimal alignment length for $s_1$ and $s_2$ are unknown. Therefore, it is advisable to initially allocate a larger number of qubits to ensure complete coverage. In this experiment, we selected a value of $L=10$ to mitigate qubit redundancy. As depicted in Fig.\ref{fig:3}(D), the resulting state is a superposition of optimal and feasible solutions, even when executed under different initial conditions. In this example, the suboptimal solution differs from the optimal solution by only a one-letter mismatch, necessitating an additional step to distinguish between them. The coexistence of these solutions further underscores the important role of quantum entanglement in solving MSA problems.\\

\subsection{Experimental Results On HYQ-A37}
Among various quantum computing devices, trapped ions, with their long coherence times and high operational fidelity, excel in precision tasks. This advantages arises from the intrinsic isolation of trapped ions from environmental noise and their manipulation using precise laser or microwave control. Moreover, ion-trap systems exhibit exceptional qubit connectivity, as ions can interact through collective vibrational modes, allowing for efficient implementation of complex quantum gates. 

\begin{figure*}[tbp]
    \includegraphics[width=\textwidth]{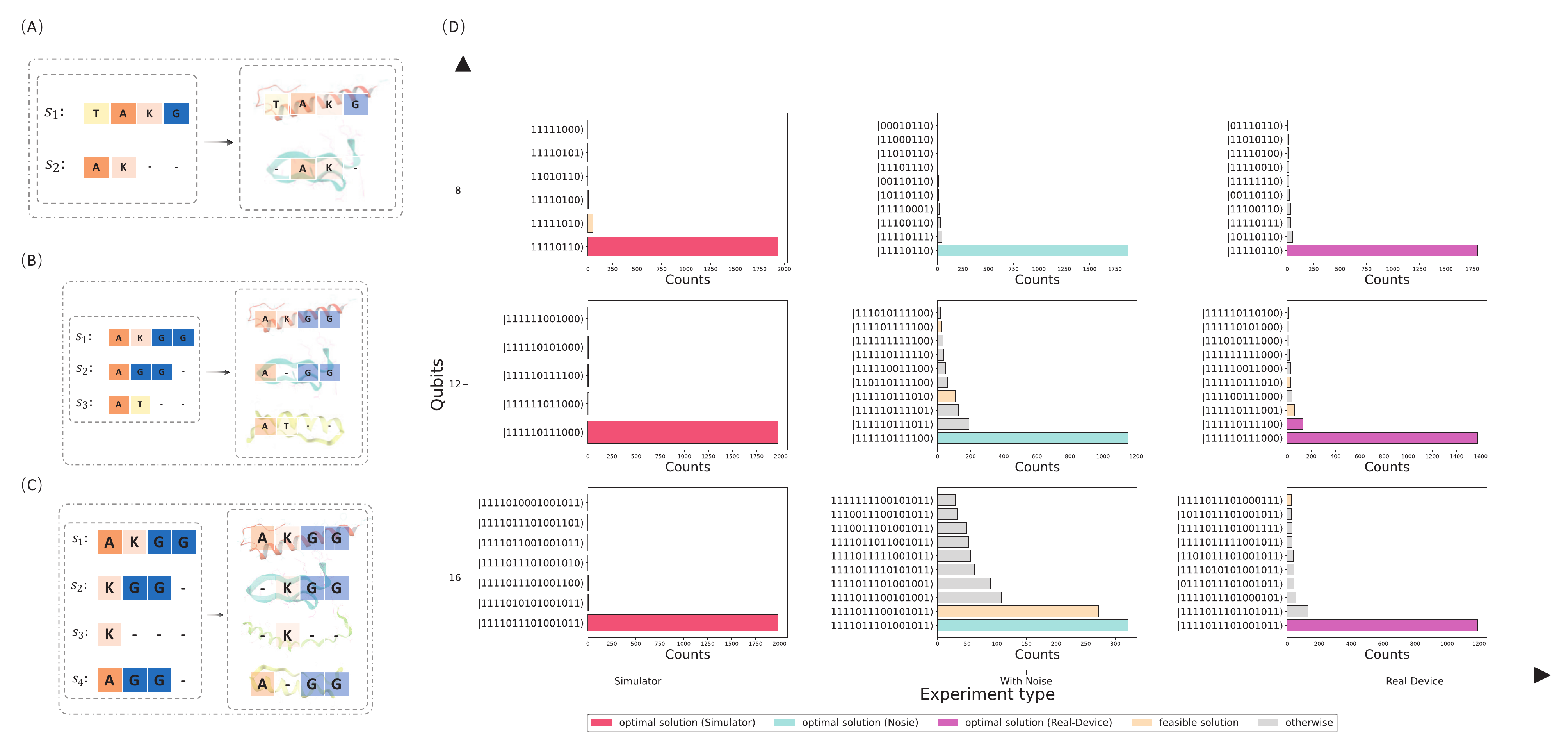}
    \caption{\textbf{The comparison of performance in three cases: simulations without noises, simulations with noises, and experiments on HYQ-A37.} (\textbf{A-C}) The example sequences are demonstrated in scenarios involving 8 , 12 and 16 qubits, respectively.  (\textbf{D}) The simulations without noises on CPU processor (indicated in red color), which exhibit the best performance in three scenatios. The simulations with noises on CPU processor (visualized in blue and yellow) that give the worst performance in all cases. The experiments carried on HYQ-A37 device (indicated in purple) showing a satisfying distribution. Note that in the 12- and 16-qubit tasks, only top 10 states are shown due to the large number of quantum states measured.}
    \label{fig:4}
\end{figure*}

In this study, our MSA model was validated on the ion-trap quantum computer HYQ-A37 (see Appendix C for details), while simulations with and without noise were executed on a CPU (AMD EPYC 9334 32-Core Processor). To make full use of limited quantum resources and alleviate the inevitable impact of quantum noise, we employed the Simultaneous Perturbation Stochastic Approximation (SPSA) algorithm for gradient estimation. Additionally, a CVaR scheme with a ratio of $r=0.8$ was employed for the 12- and 16-qubit experiments on HYQ-A37 to effectively minimize the number of optimization steps. The results from three scenarios—featuring 8, 12, and 16 qubits—were compared in Fig.\ref{fig:4}. Based on 2000 shots measurement, the noise-free simulation exhibited the best performance, benefiting from an idealized computational environment. In contrast, while the experimental results showed some susceptibility to noise and external disturbances, they still demonstrated a favorable probability distribution. Ultimately, the classical simulation incorporating noise yielded the least favorable results, likely due to an inadequate noise model as system size increases. These results demonstrated the robustness and compatibility of our approach on current quantum devices for identifying optimal solutions, even in the absence of error mitigation.    

\begin{figure*}[tbp]
    \centering
    \includegraphics[width=\textwidth]{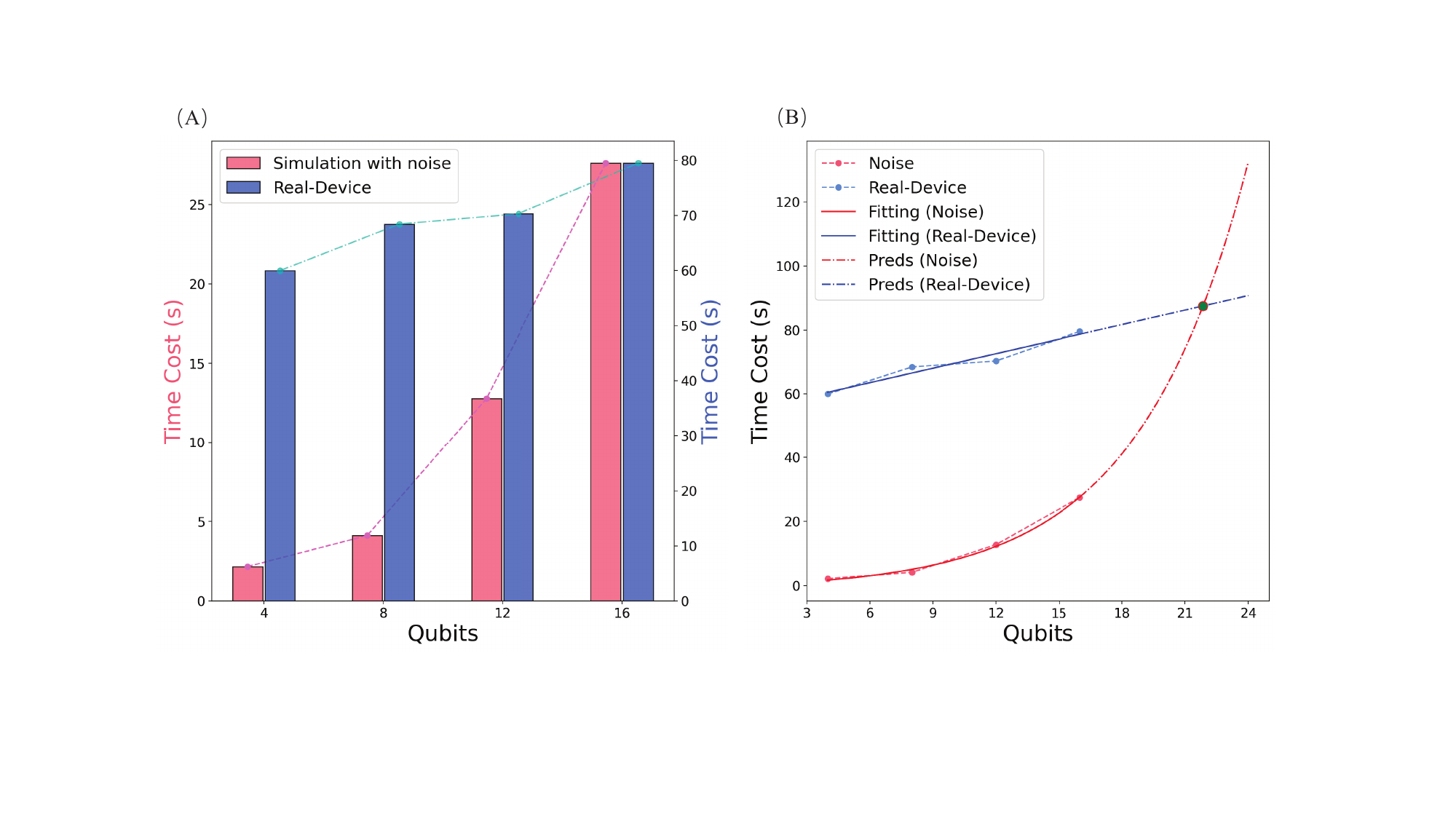}
    \caption{\textbf{The comparison of time efficiency shows classical simulations, represented in red on the left, and experimental operations on the real device, depicted in blue on the right.} \textbf{(A)} The average time required for each iteration step to reach convergence at 4, 8, 12 and 16 qubits. \textbf{(B)} The fitting and perdicting curves of time efficiency in two situations. The time cost of experiments on quantum device follows an approximately linear trend, as indicated by the blue line, whereas simulations with noise exhibit an almost exponential increase, as depicted by the red line.}
    \label{fig:5}
\end{figure*}

Furthermore, we provided an estimation of time costs based on both classical simulations and experimental data to compare current efficiency, as shown in Fig.\ref{fig:5}.  This figure displays the estimated average time required for each iteration step across all scenarios. We observed that iteration time on HYQ-A37 grows polynomially, whereas classical simulations exhibit an exponential increase. This polynomial trend can be attributed to two factors. First, the precision of the current HYQ-A37 relies heavily on real-time monitoring and irregular calibration, which account for over 50 percent of the overall time overhead and scale linearly with the number of iteration steps. Second, the gates per layer require only a few milliseconds, so an increase in circuit depth incurs at most a polynomial cost. According to our rough estimation, a crossover point arises around 22 qubits, beyond which executing our algorithm on HYQ-A37 becomes advantageous. Note that we do not account for the number of iteration steps required for convergence as qubit size increases; thus, this estimation is purely qualitative rather than exact. While scaling up the number of logical qubits in real quantum machines remains a significant challenge, classical computers still maintain a clear advantage for small- to medium-scale simulations. However, as quantum technologies continue to evolve, they are expected to surpass classical computers in solving more complex problems.

\section{Discussion}\label{sec3}
To achieve more accurate and efficient protein structure prediction, overcoming the NP-hard nature of MSA remains a core challenge. As demonstrated by AlphaFold's significant success in high-accuray protein structure predictions, MSA leverages coevolutionary features as its primary input to drive its predictions. However, the computational complexity of MSA limits scalability, particularly for large datasets. Addressing this bottleneck requires innovative solutions, such as advanced algorithms or higher performance computing technologies, to enhance the precision and efficiency of models like AlphaFold, RoseTTaFold\cite{baek2021accurate}, ESMFold\cite{lin2023evolutionary} and \textit{etc.} Quantum computing holds significant promise for addressing complex optimization problems, including MSA as a cornerstone task in computational biology. Compared with classical algorithms for MSA, quantum computing, with its abilty to explore vast solution spaces in parallel, offers a transformative approach to tackling such challenges. \\

In this study, we have developed a hybrid quantum-classical algorithm to deal with MSA tasks, which was simulated on classical computer and validated on an ion trap device HYQ-A37. We firstly mapped the MSA task to a hqQUBO model and then employed a hardware efficient algorithm to solve it. We also introduced a novel hybrid-query encoding scheme to minimize redundancy, enabling the efficient allocation of qubits on HYQ-A37 for various scenarios. Furthermore, we conducted a comparative analysis among the noise-free model, the noise-induced model, and the experimental results, all of which consistently produced optimal and feasible solutions.  Based on 2000-shot measurements, our experimental results indicated that as the number of qubits increases, the system reliably generated optimal solutions, with a significant proportion of outcomes being optimal even in the absence of error mitigation. Additionally, we assessed the average time cost per iteration in both simulations and experiments. Acccording to the fitting curves, the time per iteration of noise-induced curve exhibited an approximately exponential growth, while that of experiments grows polynomially, highlighting the time efficiency of quantum computing.\\

Our method represented an effective and robust quantum-classical algorithm tailored for MSA problems which is compatible with current quantum devices. It incorporated several innovative features. First, it employed a novel hqQUBO model that offloads the querying task onto a classical computer, thereby reserving the sampling process for the quantum hardware. This design balanced the computational load between quantum and classical systems, significantly reducing the quantum resource requirements for a given task. Second, it relied on a hardware-efficient  ansatz comprising shallow quantum circuits with modest quantum entanglement, making it particularly suitable for the limited circuit depths of NISQ-era devices and helping to avoid barren plateaus. Finally, it was optimized through a two-stage CVaR scheme that further decreases the number of iteration steps needed for convergence and mitigates the risk of becoming trapped in local minima.\\

In contrast with MSA tasks in practical applications, our model focused on a collection of short sequence alignments, serving as a proof of concept to demonstrate the effectiveness of our method and the capabilities of current quantum computers. Extending our model to address more intricate problems remains a promising direction.  As one possible application extension, a model describing DNA sequence reconstruction was given in Appendix B. To conclude, we can envision quantum or quantum-classical algorithms that leverage the inherent parallelism of qubits, enabling them to tackle complex combinatorial problems with unprecedented speed and precision. In the long run, quantum computing may revolutionize the way we approach these computationally intensive problems, driving breakthroughs in life sciences, and beyond.

\section{Method}\label{sec4}
Quantum computers offer advantages over their classical counterparts for specific problem domains. However, effectively loading data and establishing a suitable quantum framework remain challenging. In this section, we first formulated the MSA problem using the SP-score, then outlined the encoding mechanism and hqQUBO model, and finally presented the framework of the hardware-efficient quantum algorithm. \\

Suppose there are multiple sequences $S = \{s_i\}_{i=1}^{N}$,  each is composed of elements from the alphabet $\Lambda = \{$A, R, N, D, C, Q, E, G, H , I , L, K, M, F, P, S, T, W, Y, V $\}$. For example $s_i = AKGG$. Let $s_{i,k} \in \Lambda$ denotes the $k$-th letter in sequence $s_i$. Define $N$ as the total number of sequences, and $L = \max\{l_i\}_{i=1}^{N}$, where $l_i $ denotes the length of sequence $s_i$.

\subsection{Hybrid Query Encoding}\label{sec3.1}
\begin{figure*}[tbp]
    \centering
    \includegraphics[width=\textwidth]{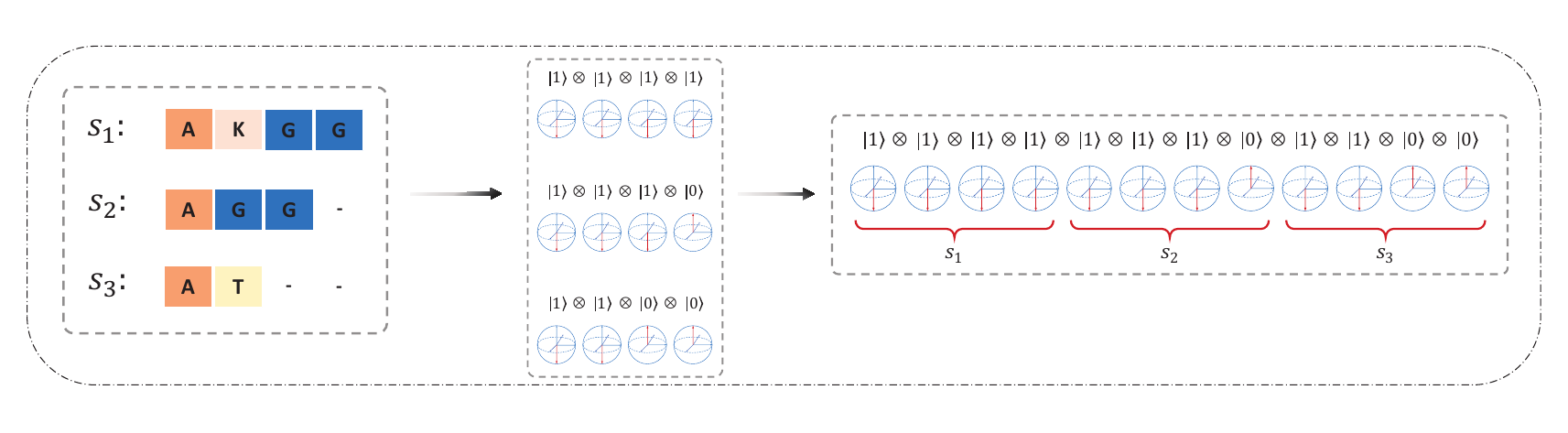}
    \caption{An example showing the hybrid querying encoding. At position $k$, the quantum state is encoded as $\ket{1}$ if there is a letter, otherwise it is $\ket{0}$.}
    \label{fig:6}
\end{figure*}

Data encoding means converting data into a representation of quantum states such that they can be recognized by a quantum device. There are various data encoding methods that define data into qubits, for more information referring \cite{weigold2021encoding}. To characterize letter positions in each sequence and circumvent excessive qubit redundancy, we introduced a novel method called Hybrid Query Encoding (HQE). In HQE, the quantum states $\ket{1}$ and $\ket{0}$ indicate whether a letter exists at position $k$, thereby preserving the original sequence order through direct queries to the initial sequence. For example, with $L = 4$, the sequence $s=AG\_\_$ is encoded as $\ket{1100}$, whereas $s' = A\_ G \_ $ is encoded as $\ket{1010}$. An illustrative example of HQE is shown in Fig.\ref{fig:6}.\\

Suppose $s_1$ is the reference sequence in the sequence group $S$, and it can be aligned with all other sequences. The number of qubits required to encode a single sequence is determined by the maximum sequence length, $L$, so HQE requires a total of $n = O(NL)$ qubits. It is important to note that HQE is a highly efficient encoding method for the MSA task, as it helps avoid redundancy. Excessive redundancy can constrain the search to feasible subspaces, as illustrated by the one-hot column encoding in this work \cite{madsen2023multi}.

\subsection{hqQUBO Formalism}
\subsubsection{Scoring Scheme}
Searching for an optimal alignment of $N$ sequences is computationally intractable as $N$ increases. Therefore, a good scoring scheme plays a very important role in evaluating all possibilities of inserting gaps into the aligned sequences. There are a couple of scoring strategies, among which SP-score (The-Sum-of-Pairs) is widely used for MSA task. The SP-score is the sum of pairwise similarity scores between all sequence pairs, which is defined as,
\begin{equation}
    SP(S') = \sum_{i \neq j} \sum_{k} \text{sim}(s'_{i,k}, s'_{j,k}), \quad \text{for} \ i, j \in N 
\end{equation}
with 
\begin{equation}
    \text{sim}(s'_{i,k}, s'_{j,k}) = \left\{
    \begin{array}{ll}
        -1, & s'_{i,k} = s'_{j,k}, \\
        +1, & s'_{i,k} \neq s'_{j,k}, \\
        \quad 0, & s'_{i,k}\  \text{or}\  s'_{j,k} = \_,
    \end{array} \quad \text{for} \ i, j \in N 
    \right.
\end{equation}
where $S' = \{ s'_{i}\}_{i}^{N}$ denotes the alignment outcome resulting from performing MSA on the original multi-sequence set $S$.\\

A dictionary $\mathcal{T}$, consisting of a series of weight matrices $w^{(i,j)}$, will be employed for querying. It serves as part of the preprocessing routine to obtain the SP-score for the HQE. Specifically, the weight matrix $w^{(i,j)} \in \mathbb{R}^{l_i \times l_j}$, and the elements of the matrix are defined as $w^{(i,j)}_{k,l} = \text{sim}(s_{i,k}, s_{j,l})$, where $\text{sim}$ denotes the similarity measure between the $k$-th letter of sequence $s_i$ and the $l$-th letter of sequence $s_j$. The SP-score can then be computed as follows,

\begin{equation}
    \sum_{i \neq j} \sum_{k} w^{(i,j)}_{f_{i}(k), f_{j}(k)} x_{i, k} x_{j, k},
\label{eq:score}
\end{equation}
where the function $f_{i}$ maps the state $\ket{1}$ in $s_{i}$ to the letter order in initial sequence, and $\ket{0}$ to $-1$, serving as a dummy index with a value of 0 upon querying. To prevent the number of letters in the final sequence from exceeding the original sequence, the corresponding state $\ket{1}$ is also mapped to $-1$. e.g., given sequence $s_{i} = AG$, if $x_{i} = \ket{0111}$, then $f_{i}(x_{i}) = [-1,0,1,-1]$. Mathematically,
\begin{equation}
    f_{i}(k) = \left\{
        \begin{array}{cc}
            -1, & \left(x_{i,k} = 0\right) \vee \left(\sum_{j=1}^{k}x_{i,j} > l_i \right)\\
            \sum_{j=1}^{k}x_{i,j}-1, & \left(x_{i,k} = 1\right) \land \left(\sum_{j=1}^{k}x_{i,j} \leq l_i \right)
        \end{array}.
    \right.
\end{equation}

\subsubsection{Constraints}
In addition, to ensure that the number of letters is conserved, we introduced the following soft constraints,
\begin{gather}
    \forall s_i, \, \sum_{k} x_{i,k} = l_i, \\
    \Leftrightarrow \sum_{i} \left( \sum_{k} x_{i,k} -l_i \right) = 0.
\end{gather}

Further, to facilitate the construction of the loss function for optimization, we converted this hard constraint into the corresponding soft constraint, i.e.
\begin{equation}
    p \sum_{i} \left(\sum_{k} x_{i,k} - l_{i}\right)^2,
\end{equation}
where $p$ means the penalty parameter, whose value depends on the scale of the problem. In our experiments, unless otherwise specified, we set the penalty parameter $p$ to 1.5, thereby prioritizing the conservation of letter count over matching errors.\\

In summary, based on HQE, we constructed the final score function as follows,
\begin{equation}
\begin{split}
    L(x; p) = &\sum_{i \neq j} \sum_{k} w^{(i,j)}_{f_{i}(k), f_{j}(k)} x_{i, k} x_{j, k} \\
                  &+ p \sum_{i} \left(\sum_{k} x_{i,k} - l_{i}\right)^2.
\end{split}
\end{equation}
Although the above formula is very close to the QUBO model, it can only be called hqQUBO because of the existence of the querying step, which is an algorithm that combines quantum and classical.

\subsection{Hamiltonian}
Once we have the score function, we can naturally derive the Hamiltonian $H$ to solve the problem. By selecting a set of bases $\{\ket{e}\}$ in the Hilbert space $\mathcal{H}$, the eigenvalues of the Hamiltonian corresponding to the MSAs of the groups of bases can be calculated by the scoring function, i.e
\begin{gather}
    H_p \ket{e} = L(\ket{e}; p) \ket{e}, \\
    \Leftrightarrow H_p = \sum_{\ket{e}} L(\ket{e}; p) \cdot \ket{e} \bra{e}.
\end{gather}
Then, the problem converts into the Variational Quantum Eigensolver (VQE) problem.

\subsection{Variational Quantum Algorithms}
VQE is one of the earliest proposed Variational Quantum Algorithms (VQAs). It utilizes a classical optimizer to train a parameterized quantum circuit (PQC) $U(\theta)$, which solves for the eigenvalues and eigenvectors of a matrix, .i.e.,
\begin{gather}
    \ket{x^{*}} = U(\theta^{*}) \ket{0}, \\
    \theta^{*} = \mathop{\arg\min}\limits_{\theta}C(\ket{x}; p), \\
    C(\ket{x}; p) = \bra{x} H_p \ket{x} = \bra{x}U^{\dagger}(\theta)H_p U(\theta)\ket{x},
\end{gather}
where $\ket{x^{*}}$ means the eigenvector with respect to the minimum eigenvalue of $H_p$, $\theta^{*}$ is the optimal parameters of PQC $U(\theta)$ corresponding to the minimum of the function $C(\ket{x}; p)$.\\

VQA is a quantum-classical hybrid algorithm based on variational optimization, whose typical workflow is similar to that of deep learning, with a key distinction: the neural network model used in deep learning is substituted with a parameterized quantum circuit $U(\theta)$. This quantum circuit minimizes the loss function $L(\left|x \right>)$ by measuring the wavefunction of its output, represented as $\left|x \right> = U(\theta)\left|0 \right>$. The optimization process is facilitated by a classical optimization solver, such as Adaptive Moment Estimation, Stochastic gradient descent, and \textit{etc}.\\

In this problem, the loss function $L(\ket{e}; p)$ is transformed into an equivalent Hamiltonian $H_p$, effectively converting the VQA problem into VQE. However, it's worth noting that in practical experiments, converting it into VQE is not a necessity. This transformation is merely illustrative to demonstrate the equivalence between the two approaches for MSA.

\subsection{Quenched CVaR Optimization Scheme}
CVaR defines a loss function based on the average over the tail of a distribution, tailored by a ratio $r$,  given by
\begin{equation}
L(\theta)_r = \langle \psi(\theta) | H_p | \psi(\theta) \rangle_r
\end{equation}
where $|\psi(\theta)\rangle_r$ is the state $|\psi(\theta)\rangle$  projected onto the MSA states subspace, consisting of states with energy in the lowest $r$-portion of the distribution.  For $r=1$,   the CVaR loss function reduces to the standard VQE loss function.\\

\begin{figure*}[tbp]
    \centering
    \includegraphics[width=\textwidth]{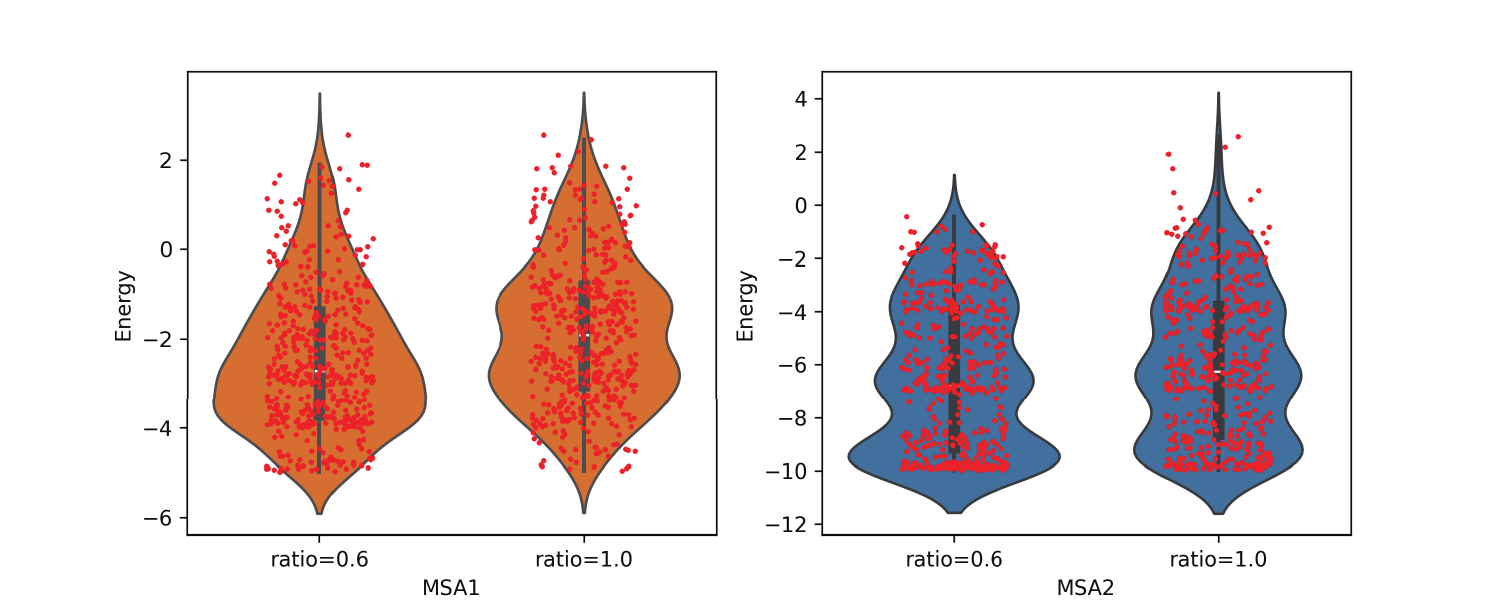}
    \caption{ The violin plot of the energy distribution for different optimization strategy in solving two MSA problems. The ratios $r=0.6$ and $r=1.0$ correspond to the loss functions used in the warm-up stage.  MSA1 (MSA2) in the left (right) subfigure corresponds to the sequencing problem in Fig. \ref{fig:3}A (C). }
    \label{fig:7}
\end{figure*}

 In our experiments, we employed the CVaR loss function and divided the optimization into two stages. First, we performed a few iterations using  $L(\theta)_{r=r_0<1}$ loss function for a warm up, then switched to the $L(\theta)_{r=1}$ loss function until convergence. Both experimental and numerical results indicate that this approach reduces the number of optimization steps and lowers the likelihood of becoming trapped in local minima. As illustrated in Fig.\ref{fig:7}, we conducted numerical simulations for both tasks in Fig.\ref{fig:3}.  During the warm-up stage,  we optimized the model using $L(\theta)_{r=r_0}$ loss function for 100 iterations, which was followed by the standard loss function for 300 iterations in the second stage. To mitigate the effect of the initial random states, the experiment was repeated 500 times. For both MSA problems, we observed a clear decrease in average energy under the two-stage CVaR optimization scheme compared to the standard approach. The SPSA algorithm was used for gradient estimation based on 2000-shot measurements.

\subsection{Hardware Platform}
All reported experimental results were performed on a trapped-ion quantum computer (HYQ-A37), details can be found in Appendix C.\\  

The HYQ-A37 is a 37-qubit ion-trap quantum computer engineered for high-fidelity quantum computing and simulation. It achieves an average error rate for state preparations and measurements (SPAM) below $0.05\%$, facilitated by electron shelving detection \cite{yang2022realizing,edmunds2021scalable}. The average fidelity exceeds $99.9\%$ and $98.5\%$ for sing-qubit and two-qubit gates respectively. The HYQ-A37 is optimized for efficient variational quantum computing, with data processing and sequence generation executed on an FPGA, and gradient descent computations performed on a classical PC. Additionally, the HYQ-A37 can be tailored to provide the requisite number of qubits and optimize the compiler to suit the specific needs of various quantum computing tasks.\\

The HYQ-A37 quantum computer is commercially available through the HYQ Cloud Services\footnote{HYQ Quantum Cloud Services: \url{https://www.q-coding.hyqubit.com}}. For the experiments described in this paper, the code and custom compiler were applied. Further information can be found on GitHub\footnote{Hardware Efficient Algorithm for Multiple Sequence Alignment on HYQ-A37: \url{https://www.GitHub.xxx.com}}.

\bigskip

\textbf{Data Availability:} The data that support the findings of this study are available from the authors upon request.

\textbf{Code availability:} The model is written in \textcolor{blue}{PyTorch v2.1.2} and \textcolor{blue}{Torchquantum v0.1.7} \cite{hanruiwang2022quantumnas}. The source code is available on GitHub (\textcolor{blue}{https://github.com}).

\textbf{Acknowledgements:} We would like to appreciate the technical support of Huayi Boao (Beijing) Quantum Technology Co., Ltd.

\textbf{Competing interests:} The authors declare that there are no competing interests.

\bibliographystyle{unsrt}
\bibliography{bibliography}

\end{document}